\documentclass[%
 reprint,
superscriptaddress,
 amsmath,amssymb,
 aps,
 prx,
]{revtex4-2}

\usepackage{graphicx}
\usepackage{dcolumn}
\usepackage{bm}
\usepackage{physics}
\usepackage{color}
\usepackage{ulem}
\usepackage{appendix}

\begin{document}

\preprint{APS/123-QED}

\title{Feasibility study on ground-state cooling and \\ single-phonon readout of trapped electrons using hybrid quantum systems}

\author{Alto Osada}
\email{alto@g.ecc.u-tokyo.ac.jp}
\affiliation{Komaba Institute for Science (KIS), The University of Tokyo, Meguro-ku, Tokyo, 153-8902, Japan }
 \affiliation{PRESTO, Japan Science and Technology Agency, Kawaguchi-shi, Saitama 332-0012, Japan}%

\author{Kento Taniguchi}
\affiliation{Komaba Institute for Science (KIS), The University of Tokyo, Meguro-ku, Tokyo, 153-8902, Japan }
 
 \author{Masato Shigefuji}
\affiliation{Komaba Institute for Science (KIS), The University of Tokyo, Meguro-ku, Tokyo, 153-8902, Japan }

\author{Atsushi Noguchi}
\email{u-atsushi@g.ecc.u-tokyo.ac.jp}
\affiliation{Komaba Institute for Science (KIS), The University of Tokyo, Meguro-ku, Tokyo, 153-8902, Japan }
\affiliation{
RIKEN Center for Quantum Computing (RQC), RIKEN, Wako-shi, Saitama 351-0198, Japan
}
\affiliation{Inamori Research Institute for Science (InaRIS), Kyoto-shi, Kyoto, 600-8411, Japan}

\date{\today}

\begin{abstract}
Qubits of long coherence time and fast quantum operations are long-sought objectives towards the realization of high-fidelity quantum operations and their applications to the quantum technologies. 
An electron levitated in a vacuum by a Paul trap is expected to be a good candidate, for its light mass and hence the high secular frequency 
which allows for the faster gate operations than those in trapped ions. 
Controlling the motional state of the trapped electron is a crucial issue, for it mediates an interaction between electron spins, intrinsic qubits embedded in electrons, and its decoherence results in degraded fidelity of two-qubit gates. 
In addition, an efficient readout of the motional state is important, regarding the possibility of detecting spin state by using it.
Despite of such an importance, how to achieve the motional ground state and how to efficiently detect it are not reported so far.  
Here we propose methods addressing these issues by utilizing hybrid quantum systems involving electron-superconducting circuit and electron-ion coupled systems and analyze the feasibility of our schemes.  
In both systems, we show that the ground-state cooling and the single-phonon readout of the motional state of the trapped electron are possible.
Our work shed light on the way to precisely control the motional states of the trapped electrons, that provides an interesting playground for the development of quantum technologies.
\end{abstract}

\maketitle


\section{Introduction}

\begin{figure}[b]
\includegraphics[width=8.4cm]{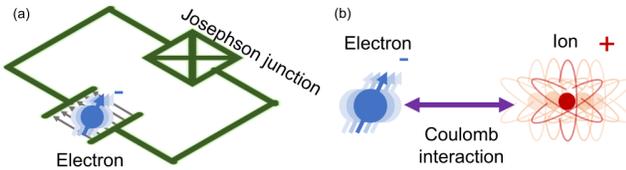}
\caption{\label{Fig1} Schematic illustrations of hybrid quantum systems considered in this work. (a) Coupled electron-superconducting qubit system where oscillatory electric field of a capacitor provides the coupling between an electron and a superconducting qubit. (b) Coupled trapped electron-trapped ion system with Coulomb attraction.}
\end{figure}

In recent development of quantum technologies, quantum bits (qubits) with long coherence times are explored in various platforms such as trapped ions~\cite{Bruzewicz2019-ac}, Rydberg atoms~\cite{Saffman2016-sn}, superconducting circuits~\cite{Kjaergaard2020-er}, and quantum defects in solids~\cite{Weber}.  For instance, hyperfine and optical qubits in atomic ions in a Paul trap possess long coherence times thanks to their intrinsic nature being protected by levitating them in ultrahigh-vacuum environment~\cite{Bruzewicz2019-ac}.  As another example, lifetimes of the superconducting qubits have improved by several orders of magnitude~\cite{Kjaergaard2020-er} for a few decades. For characterizing how good the coherence times are for various qubits, a ratio $\epsilon$ of the coherence time to the longest duration of quantum operation is useful, for its inverse limits the minimum-achievable error rates. In terms of this index, $\epsilon \sim 10^5 $ and $\epsilon \sim 10^3$ are realized so far in hyperfine qubit of atomic ions~\cite{Gaebler2016-rd} and superconducting qubits~\cite{Foxen2020-ti}, respectively. 
To further develop error-corrected qubits by e.g., using a surface code~\cite{Kitaev2003-so, Bravyi1998-jl}, a large number of physical qubits should form a single logical qubit. In this regard, it is highly demanded to achieve higher value of $\epsilon$ since it is beneficial for achieving higher fidelities and thus for reducing the required number of physical qubits for a single logical qubit. 
\begin{figure*}[t]
\includegraphics[width=17cm]{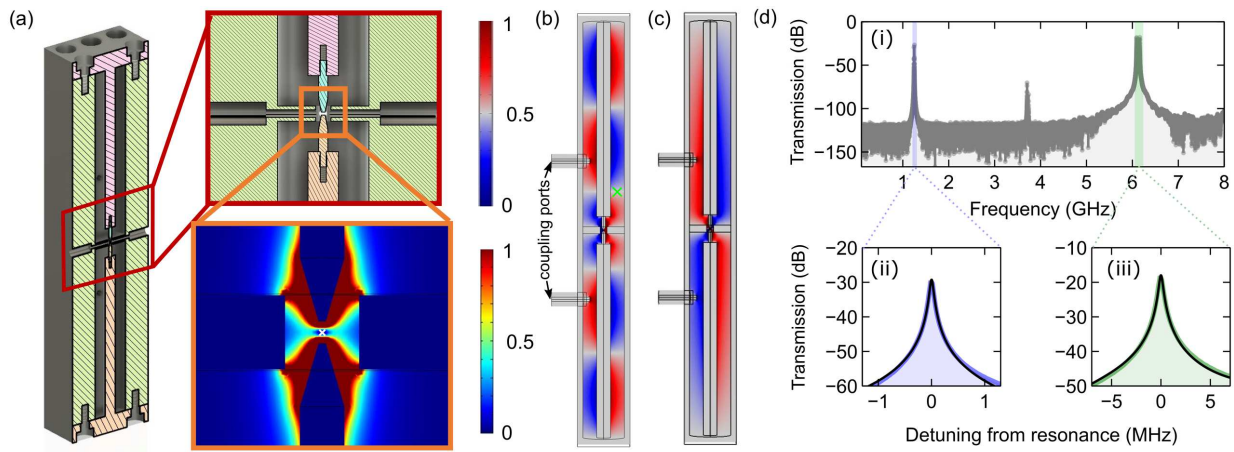}
\caption{\label{Fig2} Microwave-trap design and its characterization. 
(a) Cross-sectional view of the coupled coaxial cavity.  Insets show the expanded view and the normalized pseudo-potential for an electron with colorscale given in the right panel of (c).
(b) Field profiles of $5\lambda/4$ mode, or trap mode, used for the formation of pseudo-potential. Green cross schematically indicates a node of the mode where a transmon qubit is placed, see Sec.~\ref{transmon_readout}.
(c) That of $\lambda/4$ mode, or the readout mode, utilized for the motional-state readout.
(d) Observed transmission spectra of the coupled coaxial cavity by two-port measurement, with expanded plots of the trap mode (indicated by green) and the readout mode (indicated by blue). The black solid curves are Lorentzian fittings. Each mode exhibits, indeed, two peaks corresponding to symmetric and anti-symmetric modes. The expanded plots display the anti-symmetric $\lambda/4$- and symmetric $5\lambda/4$ modes under concern.
}
\end{figure*}

For such a purpose, trapped-electron system~\cite{Wineland1973-se,Daniilidis_2013,Kotler_2017,Matthiesen2021-mf,Yu2022-zz} is a good candidate. An electron is a spin-1/2 system, literally being a pure two-level system or an intrinsic qubit without any other internal states into which the quantum information leaks. ikewise in the trapped-ion electron-spin qubit~\cite{Ruster2016-jp}, the levitated electron-spin qubit is expected to maintain its coherence for orders-of-magnitude longer time compared to the solid-state counterparts [see e.g., Ref.~\cite{Zwerver2022-xl} and references therein]. Trapped electrons also exhibit phonon degrees of freedom as trapped ions do, in which the phonon can be used as a quantum bus to implement a two-qubit gate~\cite{Molmer1999-dq} and can be utilized for detecting the spin state~\cite{Peng_2017}. It is noteworthy that owing to more than $10\,000$ times lighter mass of an electron than those of ions, frequency of the phonon mode can be made as high as several hundreds of MHz or even be designed in the GHz region. Therefore, two-qubit gates in the trapped electron system are expected to be implemented within tens of nanoseconds when limited by the oscillation period of the electron. On top of that, zero-point fluctuation $x_{\mathrm{zpf}}$ of the phonon reaches 100~nm range, 10 times greater than the ions'.  The larger the zero-point fluctuation is, the stronger the interaction to the other systems exhibiting electromagnetic waves.  In short, trapped electrons potentially possess longer coherence time than superconducting qubits and shorter gate duration than trapped ions, and thus are promising building blocks of the quantum technologies.

Electrons in a Paul trap have just been realized recently~\cite{Matthiesen2021-mf}, and high-fidelity quantum operations~\cite{Yu2022-zz}, spin-state detection~\cite{Peng_2017} and coupling to superconducting electronics for the quantum-state conversion~\cite{Daniilidis_2013} are discussed.  A few more things should be addressed in order to fully utilize above advantages. First thing is a highly sensitive readout of the motional state of a single trapped electron. 
The readout of motional state, in particular, is of great interest in the scope of phonon-mediated spin-state readout as discussed in Ref.~\cite{Peng_2017}.
Second, electrons' motion has to be cooled down close to its vibrational ground state to eliminate the gate errors. Resistive cooling~\cite{Itano_1995, Wineland1973-se} at low temperature will help electrons to get sympathetically cooled with the trapping electrodes, that said, active methods of cooling electrons are preferable since such cooling process does not afford the ground-state cooling at 4~K. We would like to mention that a levitated electron does not have any optical transitions, that hinders the straightforward application to quantum network~\cite{Kimble2008-bc}. Therefore, the connectivity of an electron to an optical photon is also a major concern.

In this article, we propose several methods that potentially resolve above issues by considering hybrid quantum systems comprised of trapped electrons, superconducting circuits and trapped ions, where electric fields of a superconducting circuit and a trapped ion can be utilized as sources of interaction (see Fig.~\ref{Fig1}). By utilizing such quantum systems, we reveal that the single-phonon readout and ground-state cooling of the motional state of electrons, and that the quantum operations are feasible not only by radio-frequency or microwave fields but also by optical means, nevertheless a trapped-electron does not have optical transition.  
We also experimentally implement a few essential tools indispensable for realizing the ideas including a superconducting resonator and a cryocooler-compatible, low-energy electron source. 
Our results add fundamental building blocks for a growing field of trapped electron and pave the way for realizing an electron-based quantum computation.

This article is constructed as follows. 
In Sec.~\ref{sec_es}, we first introduce a coupled coaxial cavity for trapping electrons and for reading out the motional state of the trapped electron, using resonant modes held in the cavity. Further using a dispersively-coupled cavity-transmon system, the electron's motion can be cooled down to its ground state and read out at the single-phonon level.
Then we switch the topic to the electron-ion coupled system in Sec.~\ref{sec_ei}.  Simultaneous Paul trapping of electrons and ions is proposed first and their coupling through the Coulomb attraction is discussed. Single-phonon readout and ground-state cooling for the trapped electron is shown to be feasible with the trapped ion as well.
Sec.~\ref{conclusion} summarizes the results.

\section{Trapped electron and superconducting circuit}\label{sec_es}

\subsection{Coaxial-cavity trap}\label{coaxcav}

By combining static and oscillatory quadrupolar electric fields, the Paul trap provides a powerful tool to trap charged particles, which has become a core technology in trapped-ion-based quantum computing~\cite{Bruzewicz2019-ac} and quantum metrology~\cite{Brewer2019-jh}.
Electrons trapped in a Paul trap exhibit the secular oscillation at the frequency of the order of GHz~\cite{Yu2022-zz}, which can be resonantly coupled to electromagnetic fields of the ultralow-loss superconducting circuits.
Superconducting coaxial cavity is utilized for cavity quantum electrodynamics (QED) and microwave quantum optics (see e.g. Ref.~\cite{Reinhold_2020} and references therein) for its high quality factor and large field concentration.  We adopt it for the electron trapping as well, as detailed below.


Figure~\ref{Fig2}(a) schematically shows a coaxial-cavity based electron Paul trap investigated in this work.
Two coaxial cavities, each consisting of a coaxial line terminated at a tip, are opposed to each other. Since each coaxial cavity forms a $\lambda/4$ resonator with grounded and floated edges, it supports a fundamental $\lambda/4$ mode and a series of higher-order modes with wavelengths assigned by $(2n+1)\lambda/4$ with $n$ being a positive integer.  The electrode tips with 400~$\mathrm{\mu m}$-diameter circular apices are placed 400 $\mathrm{\mu m}$ apart from each other, which yield the distance between the effective-potential minimum and the electrode tip $r_0$ to be 200 $\mathrm{\mu m}$. The upper and lower coaxial cavities couple to each other capacitively by the electrode tips to result in the appearance of hybridized modes with symmetric and anti-symmetric electric-field distributions.
The symmetric mode exhibits a quadrupolar electric field between the electrode tips to form an effective potential for the electron. The effective-potential landscape is shown in the right-bottom inset of Fig.~\ref{Fig2}(a) which is obtained by numerical method using COMSOL.
The whole electric-field distribution of the symmetric $5\lambda/4$ resonance, which is called trap mode here, is shown in Fig.~\ref{Fig2}(b) as well. 
The simulated trap mode possesses the resonant frequency of $6.0$~GHz. By assuming that its quality factor amounts $10^6$~\cite{Reinhold_2020}, the effective potential yields $3.0$~eV depth and the secular frequency of $1.2$~GHz with $0.8$~mW input.

In contrast, the anti-symmetric mode generates a dipolar field at the effective-potential minimum.  This can be utilized for a coupling to a secular motion of electrons and allows us for cavity-assisted electrical detection of them, as will be discussed further in Sec.~\ref{sec_es_read}. We utilize an anti-symmetric $\lambda/4$ resonance, which we call readout mode, of the same coupled coaxial cavity for this purpose, since it exhibits the resonance frequency of $1.2$~GHz that is resonant on the secular motion of the electrons in the design described above. The electric-field distribution of the readout mode is displayed in Fig.~\ref{Fig2}(c). 
 

In order to investigate the realizability of above trap design, we constructed and evaluated the coupled coaxial cavity made of aluminum in practice under 300~mK environment realized by a $^3$He-based cryocooler.  We implement two-port measurements through two coupling ports attached symmetrically to the two coaxial cavities.  The transmission spectra are shown in Fig.~\ref{Fig2}(d). In the spectrum Fig.~\ref{Fig2}(d-i), three peaks at 1.2~GHz, 3.7~GHz and 6.1~GHz correspond respectively to $\lambda/4$, $3\lambda/4$ and $5\lambda/4$ resonances.  The spectra of readout and trap modes are plotted in Fig.~\ref{Fig2}(d-ii) and (d-iii), respectively.  Black curves are Lorentzian fittings that yield the intrinsic quality factor of $1.8\times10^4$ for the readout mode and a similar value is obtained for the trap mode.
Even with this quality factor of the trap mode being lower than the assumed value of $10^6$ in the numerical estimation, $2$~mW injection is estimated to give the same effective potential for the electrons if we use optimized, sharper electrode tips, without spoiling the cryogenic environments.


\subsection{Cryocooler-compatible low-energy electron source} \label{ablation}


In the recent room-temperature implementation of Paul trapping of electrons~\cite{Matthiesen2021-mf}, neutral atoms from an oven are photoionized to realize the \textit{in situ} generation of electrons of sufficiently low energy for the trapping. However, cryogenic environment does not afford devices with large heat production exceeding the cooling power. Therefore, the use of oven is hindered and an alternative method is required. In such a perspective, pulsed laser ablation~\cite{Dubielzig2021-qh, Vrijsen2019-xr, Shao2018-ak, Osada2022-oh} for the generation of neutral atoms is widely recognized as a cryocooler-compatible method, for it utilizes short-term, local heating of the target material to generate a tiny jet of atom flux. 

\begin{figure}[t]
\includegraphics[width=8.6cm]{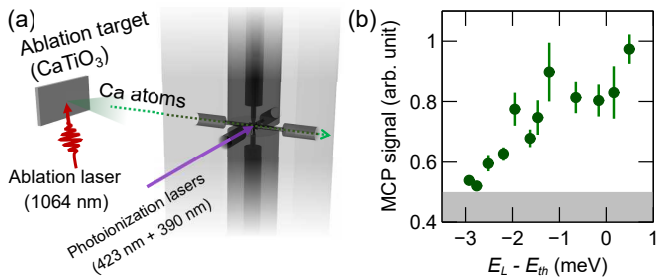}
\caption{\label{Fig3} Cryocooler-compatible electron generation. (a) Proposed experimental system of the electron generation based on Ca atom generation by 1064~nm-wavelength nanosecond pulsed-laser ablation and succeeded photoionization. In experiment (b), a multi-channel plate (MCP) is placed instead of the coupled coaxial cavity. (b) Signal of electrons detected by an MCP. The electrons are generated by the combination of pulsed-laser ablation and photoionization as depicted in (a). Horizontal axis represents the energy of the 390~nm-wavelength ionization laser, relative to the $E_\mathrm{th}$. The upper bound of the gray shaded region stands for the background signal level, namely the MCP signal without the 423~nm-wavelength photoionization laser. Error bars represent one standard deviation.}
\end{figure}

Such a method is directly applicable to the electron generation that is naturally involved in the photoionization of neutral atoms. Figure~\ref{Fig3}(a) shows a schematics of a proposed low-energy electron source. A pulsed nanosecond 1064~nm-wavelength laser impinges on an ablation target to generate a neutral atoms to be photoionized.  For instance, calcium atoms are ablated out of the calcium titanate (CaTiO$_3$), which is revealed to be a long-lived calcium atom source enduring over thousands of ablation pulses (see Appendix A). A jet of calcium atoms is then introduced into the trapping region of the coupled coaxial cavity. Other two lasers also points toward the trapping region to photoionize the atoms, by driving optical transitions of calcium atom with transition wavelengths 423~nm ($^1S_0$ $\rightarrow$ $^1P_1$) and 390~nm (from $^1P_1$ to continuum)~\cite{Matthiesen2021-mf,Gulde2001-dc}.  Generated electron carries excess energy brought by the second laser, or more concretely, the energy difference $E_{\mathrm{L}}-E_{\mathrm{th}}$ of the incident 390~nm-wavelength photon $E_{\mathrm{L}}$ and the ionization-threshold energy $E_{\mathrm{th}}$ from $^1P_1$ state, that corresponds to the wavelength of $389.81$~nm~\cite{Gulde2001-dc}, can naively be regarded as the energy of the generated electrons.  Therefore, the energy of the electron can be tuned by $E_{\mathrm{L}}$ in this scheme.  In addition, the electron generation process is comprised of the laser ablation and the photoionization, so that this all-optical method can be done in a pulsewise manner, resulting in negligible heat production and hence compatibility with a cryogenic environment.

To demonstrate our electron generation scheme, we placed a CaTiO$_3$ as an ablation target and a multi-channel plate (MCP) with electrodes for guiding and accelerating electrons in a vacuum chamber at room temperature.   In this experiment we do not have coupled coaxial cavity in the ionization region. The laser pulse for ablation is applied with its single-pulse energy of 76~$\mu$J and corresponding pulse fluence of 1~J/cm$^2$. When the repetition rate of the pulse is 1~Hz, this pulse irradiation results in only 76~$\mu$W heat generation if all the energy was absorbed by the sample. Such an experimental condition allows us to observe sizable signals of electrons at MCP. In Fig.~\ref{Fig3}(b), MCP signal, which is proportional to a number of generated electrons, is plotted for various energies $E_{\mathrm{L}}$ of the photoionization laser around the ionization threshold $E_{\mathrm{th}}$.  As $E_{\mathrm{L}}$ approaches and exceeds $E_{\mathrm{th}}$, a bright spot indicating the detection of electrons appears on the phosphor screen of the MCP. This signals the successful generation of electrons by the combination of laser ablation and succeeding photoionization of calcium atoms. We could also demonstrate the compatibility of this method with a $^3$He-based cryocooler with a separate experiment.  

Considering that the energy difference $E_{\mathrm{L}}-E_{\mathrm{th}}$ corresponds to the energy of the electron, successfully generated electrons in Fig.~\ref{Fig3}(b) are deduced to have their kinetic energy of less than several meV, which is much smaller than the depth of the effective potential.

\subsection{Readout of motional state} \label{sec_es_read}

\subsubsection{Electrical readout}

As mentioned in Sec.~\ref{coaxcav}, we utilize the anti-symmetric $\lambda/4$ mode of the coupled coaxial cavity for the readout of the secular motion of trapped electrons.  The predominant interaction between microwave field and such phonon mode of an electron, which we call it simply the phonon hereafter, is electric dipole interaction. Let the phonon mode and the readout mode respectively have their eigenfrequencies $\omega_{\mathrm{e}}$ and $\omega_{\mathrm{MW}}$, with their respective annihilation operators $\hat{a}$ and $\hat{b}$.  The phonon-cavity interaction described above leads to a total Hamiltonian $\hat{H}_{\mathrm{ec}}$ of such a hybrid system written as
\begin{equation}
    \hat{H}_\mathrm{ec} = \hbar\omega _\mathrm{e} \hat{a}^\dagger\hat{a} + \hbar \omega _\mathrm{MW} \hat{b}^\dagger\hat{b} + \hbar g_\mathrm{ec} (\hat{a}^\dagger\hat{b} + \hat{a}\hat{b}^\dagger)
\end{equation}
        where the phonon-cavity interaction is incorporated in the third term under the rotating-wave approximation. The coupling strength reads $g_{\mathrm{ec}} = e x_\mathrm{zpf} E_\mathrm{zpf}$ with $x_\mathrm{zpf}=\sqrt{\hbar/2m_\mathrm{e}\omega_\mathrm{e}}$ representing zero-point fluctuation of the phonon mode and $E_\mathrm{zpf}$ being that of the electric field of the readout mode, respectively.  The latter one, $E_\mathrm{zpf}$, depends on the circuit design and numerical simulation using COMSOL reveals that $g_{\mathrm{ec}}/2\pi = 33$~kHz in our coupled coaxial-cavity design. 

The hybrid system then manifest itself as being in a strong coupling regime with such a coupling strength, by assuming that an intrinsic quality factor of more than 100\,000 can be achieved for the readout mode.  We can utilize this hybrid system for reading out the phonon state, by extracting microwaves radiated from the electron through the microwave cavity.  The extraction will be efficient owing to the strong phonon-cavity coupling mentioned above and properly tuned external coupling of the cavity.  With the use of a commercial high-electron-mobility-transistor amplifier, the sensitivity for the oscillatory motion of an electron can be estimated to be $-195$~dBm/Hz, corresponding to that of a few tens of phonons with unity signal-to-noise ratio.

\subsubsection{Quantum non-demolition measurement using a Josephson circuit} \label{transmon_readout}

As stated above, the phonon state can be read out resonantly through the superconducting microwave cavity, however, sensitivity is still insufficient for single-phonon detection. In addition, the coupling strength $g_\mathrm{ec}$ is not enough for phonon-number resolving measurement.  With this observation, we are motivated to utilize another ingredient of a Josephson circuit, or more concretely a three-dimensional transmon qubit in our scheme.  The transmon qubit has been placed inside the coupled coaxial cavity around the antinode of the cavity mode to implement cavity QED system in strong dispersive regime~\cite{Wallraff2004-fa}.  In our cavity design, the transmon qubit should be put at the node of the trap mode [see Fig.~\ref{Fig2}(b)] so that it is unaffected by the presence of the transmon qubit.  Transmon qubit at this position can still strongly couple to the readout mode.  Let $g_\mathrm{sc}$ denote the coupling strength between the readout mode and the transmon qubit represented by Pauli operators. The Hamiltonian of the readout system now reads
\begin{equation}
    \hat{H}_\mathrm{read} = \hat{H'}_\mathrm{ec}+\hat{H}_\mathrm{sc}
\end{equation}
with $\hat{H'}_\mathrm{ec} = \hbar\omega _\mathrm{e} \hat{a}^\dagger\hat{a} + \hbar g_\mathrm{ec} (\hat{a}^\dagger\hat{b} + \hat{a}\hat{b}^\dagger)$ and $\hat{H}_\mathrm{sc}$ defined by
\begin{equation}
    \hat{H}_\mathrm{sc} =\hbar \omega_{\mathrm{MW}} \hat{b}^\dagger\hat{b} + \frac{\hbar\omega _\mathrm{q}}{2}\hat{\sigma}_z 
    +\hbar g_\mathrm{sc} (\hat{b}^\dagger\hat{\sigma}_{-}+\hat{b}\hat{\sigma}_{+} ).
\end{equation}
Here $\omega_\mathrm{q}$ is the transition frequency of the transmon qubit. 

\begin{figure}[t]
\includegraphics[width=8.6cm]{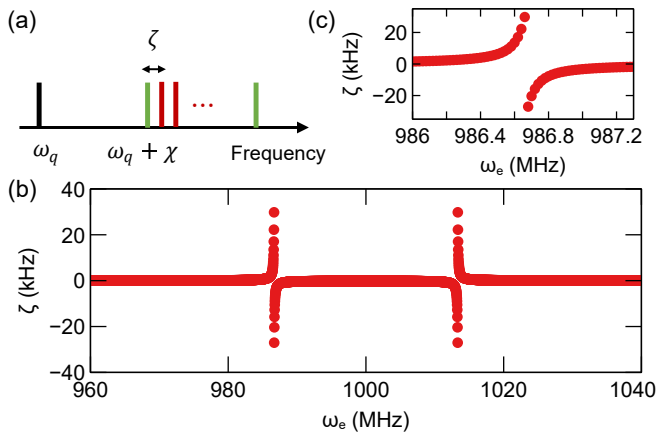}
\caption{\label{Fig4} Dispersive interaction of phonon of an electron and transmon qubit. (a) Schematic illustration of transmon-qubit spectrum with dispersive interactions. The original frequency of the transmon qubit (black) is predominatly split into multiple copies (green) by the dispersive interaction with a microwave cavity mode, where the frequency spacing is approximately given by $\chi = g_\mathrm{sc}^2/\Delta_\mathrm{sc}$.  Since the microwave cavity mode couples to the phonon mode of an electron, their spectra are dispersively shifted (indicated by red) as well, with the coupling strength $\zeta$. (b) Numerically obtained values of $\zeta$ as a function of the phonon frequency $\omega_\mathrm{e}$. Detailed  structure of the resonant behavior around $\omega_\mathrm{e}/2\pi =986$~MHz is shown in (c). Parameters used for this calculation are listed in Table.~\ref{Tab1}.}
\end{figure}

Suppose that the phonon frequency is resonant on the microwave cavity mode, while the transmon qubit is off-resonant.  The coupled phonon-cavity system exhibits two normal modes, and the transmon qubit dispersively couples to the normal modes to experience a Stark effect in the presence of the phonon of the electron motion.  Under a reasonable situation of $g_\mathrm{ec} \ll g_\mathrm{sc}$, we can first derive the dispersive Hamiltonian by applying Schrieffer-Wolff transformation~\cite{Schrieffer1966-gm} on $\hat{H}_\mathrm{sc}$ without the trapped-electron system and the coupling to it, yielding a dispersive Hamiltonian $\hat{H}'_\mathrm{sc}$ involving the readout mode and the transmon qubit~\cite{Wallraff2004-fa} as 
\begin{align}
    \hat{H}'_\mathrm{sc} = \frac{\hbar}{2}\left( \omega_\mathrm{q} + \frac{g_\mathrm{sc}^2}{\Delta_\mathrm{sc}} \right)\sigma_z + \hbar \left( \omega_\mathrm{MW} + \frac{g_\mathrm{sc}^2}{\Delta_\mathrm{sc}}\sigma_z \right) b^\dagger b
\end{align}
with $\Delta_\mathrm{sc} = \omega_\mathrm{MW} - \omega_\mathrm{q}$. Then by adding $\hat{H}'_\mathrm{ec}$ to it, further Schrieffer-Wolff transformation for $\hat{H}'_\mathrm{sc}+\hat{H}'_\mathrm{ec}$ brings the readout Hamiltonian into an approximately diagonalized form by regarding the off-diagonal term $\hbar g_\mathrm{ec} (\hat{a}^\dagger\hat{b} + \hat{a}\hat{b}^\dagger)$ as a perturbation. The latter transformation is executed symbolically by Mathematica program. The resultant effective Hamiltonian contains the dispersive coupling between phonon mode and the transmon qubit $\hbar \zeta \hat{a}^\dagger\hat{a} \sigma_z$ with the coupling parameter $\zeta$ given by
\begin{align}
    \zeta = \frac{2g_\mathrm{ec}^2g_\mathrm{sc}^2 \Delta_\mathrm{sc}}{g_\mathrm{sc}^4 - \Delta_\mathrm{ec}^2\Delta_\mathrm{sc}^2}.
\end{align}
Here the detuning $\Delta_\mathrm{ec} = \omega_\mathrm{MW} - \omega_\mathrm{e}$ is introduced. Let us define here another detuning $\delta = \Delta_\mathrm{ec}-g_\mathrm{sc}^{2}/\Delta_\mathrm{sc}$. This stands for the frequency difference between the phonon and the microwave cavity mode, where the latter one is now shifted due to the dispersive coupling to the transmon qubit. We can safely assume $|\delta| \ll |g_\mathrm{sc}^2/\Delta_\mathrm{sc}|$ to rewrite the above expression as 
\begin{align}
    \zeta = -\frac{g_\mathrm{ec}^2}{\delta}.
\end{align}
This is indeed a dispersive coupling between the phonon and the microwave cavity mode and is valid when $|\delta| \gg g_\mathrm{ec}$.  The coupling term $\hbar \zeta \hat{a}^\dagger\hat{a} \sigma_z$, after all, imprints the phonon-number-dependent spectrum of coupled phonon-cavity system onto the transmon spectrum [see Fig.~\ref{Fig4}(a)]. Even if $|\delta| \gg g_\mathrm{ec}$ does not hold, i.e. the coupled phonon-cavity system is near resonant, $|\zeta| \simeq g_\mathrm{ec}$ can be obtained at most. For comparison, we plot $\zeta$ with respect to $\omega_\mathrm{e}$ obtained by numerical diagonalization of $\hat{H}_\mathrm{read}$ in Fig.~\ref{Fig4}(b) and (c).  The parameters used for this calculation are listed in Table.~\ref{Tab1}. 

\begin{table}[b]
 \centering
  \begin{tabular}{clll}
   \hline
   microwave-cavity-mode frequency, $\omega_\mathrm{MW}$ & 1~GHz \\
   transmon-qubit frequency, $\omega_\mathrm{q}$ & 4~GHz \\
   phonon-cavity coupling strength, $g_\mathrm{ec}$ & 33~kHz \\
   qubit-cavity coupling strength, $g_\mathrm{sc}$ & 200~MHz \\
   \hline
  \end{tabular}
  \caption{ Parameters used for the evaluation of $\zeta$.}\label{Tab1}
\end{table}

In the strong dispersive regime in which $\zeta$ overwhelms decoherence rates of the phonon mode, readout mode and transmon qubit, the phonon number will be resolved, likewise in the circuit QED systems~\cite{Schuster2007-ii}. With the large quality factor $\sim 10^6$ of the cavity mode and the long coherence time $\sim 60$~$\mu$s of the qubit~\cite{Reinhold_2020}, a phonon-number resolving experiment is feasible with a reasonable assumption that the coherence time of the phonon mode is as long as $10$~ms.

\subsection{Cooling with superconducting circuits}


Phonon mode of trapped electrons can be used for implementing a two-qubit gate and its thermal excitation give rise to errors of the gate operations. Therefore, the phonon is preferably prepared in its ground state, however, electrons do not possess internal structure except for the intrinsic spin degree of freedom, so that a powerful method of laser cooling, which enables us to obtain atomic ions in their motional ground state, is unavailable for the trapped-electron system.
A trapped electron loses its motional energy predominantly through the resistive cooling~\cite{Itano_1995, Wineland1973-se} that equilibrates the electron with thermal environment.  At 300~mK environment with motional frequency of 1~GHz, mean phonon occupation number $\bar{n}$ can be roughly estimated by assuming the Bose-Einstein distribution to be $\bar{n} \simeq 6$. Moreover, it is revealed that the heating rate for trapped electrons amounts 140 quanta per second~\cite{Yu2022-zz}, which may compete the resistive-cooling rate. Other ways than just using dilution refrigerator to cool the trapped electrons are highly demanded, and in this Section, we discuss two cooling methods: one using a cavity mode and another one using a transmon qubit and its measurement.  

\subsubsection{Cooling using a microwave cavity}

A bottleneck of the resistive cooling is its cooling rate, that affords only partial cooling of electron's motion but not down close to its motional ground state, or even higher due to the competition with the heating rate.  A straightforward idea of improving the cooling rate is to utilize a microwave resonator that is resonant with the phonon mode~\cite{Kotler_2017}. An oscillatory motion of an electron can couple to the readout mode and it allows for the coupling of phonon mode to the propagating mode with the rate $\sim 33$~kHz, as discussed in the previous Section.  This cooling rate well surpasses the conservatively assessed heating rate of $140$~Hz~\cite{Yu2022-zz}. Thanks to this, the phonon mode equilibrates well with the thermal bath, which yields the mean phonon number of $6$ at $300$~mK environment.

\subsubsection{Cooling using a transmon qubit}


We have just considered the cooling of a phonon mode by the use of a coupling to the readout mode.  Even with such a method, it allows the trapped electron to thermalize with the cryogenic environment, that is, the artificially added cooling and heating rates are relatively large but still totally equal.  Further cooling beyond the temperature of the cryocooler $\sim 300$~mK is required for achieving the motional ground state.

Our approach is to utilize a circuit-QED system where a transmon qubit dispersively coupled to the readout mode.  The coupling between them is typically in the MHz regime, which is much larger than $g_\mathrm{ec}$, so that we can consider the circuit-QED system first and then take into account the trapped electron afterwards.  We consider circuit-QED system and its quantum state $\ket{\xi,n }$ with the number of excitations $n$ in the coupled coaxial cavity and the qubit state $\xi = g$, $e$ or $f$ respectively corresponding to the ground, first-excited or second-excited state. The cooling process we propose is as follows.  First we consider the state $\ket{g,1}$ and apply a $\pi$ pulse of  $\ket{g,1}\leftrightarrow\ket{f,0}$ transition to transfer the excitation from the cavity to the qubit, and consecutively apply a $\pi$ pulse of $\ket{e,0}\leftrightarrow\ket{f,0}$ transition to bring the system in $\ket{e,0}$ state. If we start with the $\ket{g,0}$ state, above procedure does nothing. Finally the transmon qubit is read out through the $3\lambda/4$ mode of the coupled coaxial cavity, which we call the auxiliary mode.  This procedure is a projective measurement thanks to the strong dispersive coupling between the auxiliary mode and the qubit, that transforms a mixed state to a pure state to reduce the entropy to result in the cooling down of the readout mode.  At this point, let us recall that the electron is coupled to the readout mode. The electron therefore thermalizes with the readout mode and thus cooled down as well.

By adopting a similar scheme, using the decay of the transmon qubit rather than a measurement of it, mean photon number in a LC resonator with its resonant frequency being $173$~MHz is cooled down to $\sim 0.1$~\cite{Gely2019-mz}, corresponding to a mode temperature of 4~mK.  
Although this value is dependent on the temperature of the environment, the scheme itself is valid for the current setup as well. By applying a $\pi$-pulse which brings $\ket{e,0}$ to $\ket{g, 0}$ instead of waiting for the qubit to decay, a single cooling cycle can be completed within 1.5~$\mu$s, given that the Rabi frequency of the $\ket{g,1}\leftrightarrow\ket{f,0}$ and $\ket{e,0}\leftrightarrow\ket{f,0}$ transitions become a few MHz and that of $\ket{g,0}\leftrightarrow\ket{e,0}$ transition is on the order of 10~MHz, with the measurement time taking $\sim$1~$\mu$s. The added cooling rate by above procedure thus amounts sub-MHz. This scheme, hence, is capable of reducing the mean photon number of the read mode down to its ground state and so for the motional state of the trapped electron.
In this manner, the cooling using the transmon qubit and its measurement provides a way to achieve motional ground state of the trapped electron.


\section{Trapped electron and trapped ion} \label{sec_ei}

In this Section, we introduce an electron-ion hybrid system in terms of the feasibility of simultaneous trapping of electrons and ions, state readout and cooling of the trapped electrons with the help of the ions.

\subsection{Simultaneous trapping of electrons and ions}

Electrons and ions interacts with each other by Coulomb interaction and thus the closer they are, the stronger the interaction becomes. Each of them can be trapped using a Paul trap, however, there are a few technical difficulties toward the simultaneous trapping of them. One is the large difference of their masses by a factor ranging from $10^4$ to $10^5$.  This results in a difference of the frequencies of their secular motions and those of supplied ac voltages to the trap, so that the design of the trap electrodes needs a special care~\cite{Leefer2016-du, Dehmelt1995-fs, Foot2018-gl}.  Second issue is their signs of the charge.  The atomic ions are positively charged and the electrons are, of course, negatively charged.  The pseudo-potential formed by ac voltages are valid for both particles, but the linear Paul trap requires dc voltages as well and a simple end-cap structure does not provide axial trapping potential for both simultaneously. Additionally, it is desirable to have electrons and ions trapped close to each other, however, it is important to trap them separately with a finite distance, in order to avoid the trapping of electrons by atomic ions that give rise to recombination processes, that is rather of interest in a different context such as quantum chemistry realized in the single-particle level.

\begin{figure}[t]
\includegraphics[width=8cm]{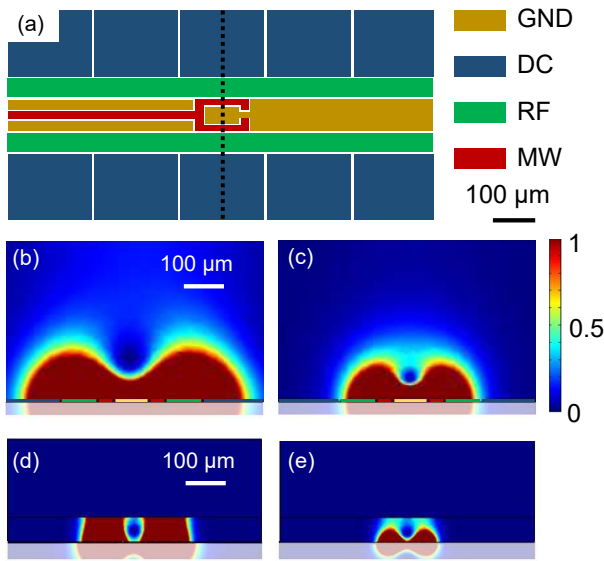}
\caption{\label{Fig5} Electrode design for simultaneous trapping of electrons and ions. (a) Surface-electrode configuration. Basic structure is a segmented five-rail linear Paul trap with a grounded electrode (GND, yellow) in the middle and RF electrodes (green) on the both sides of it.  Confinement in the axial direction is realized by the segmented dc electrodes (blue).  Grounded electrode is further divided and red part acts as a microwave electrode (MW), which can be regarded as a microwave ring trap and as a grounded one in the RF region. (b, c) Normalized pseudo-potential profiles for (b) an ion and (c) an electron in the cross-section along the black dotted line in (a). Scale bar is common for (b) and (c). (d, e) Normalized pseudo-potential profiles for (d) an ion and for (e) an electron with a grounded ceiling at 100~$\mu$m above the trap electrodes. }
\end{figure}

Our idea is to utilize a surface electrode trap with RF electrodes for trapping ions and microwave electrodes for electrons coexisting in-plane. A top view of the electrode design is schematically shown in Fig.~\ref{Fig5}(a). The electrode structure is a segmented five-rail linear Paul trap with grounded electrode (GND, yellow) in the middle and RF electrodes (green) on the both sides of it.  Confinement in the axial direction is realized by the segmented dc electrodes (blue).  The grounded electrode is further divided and red part acts as a microwave electrode (MW), which can be regarded as a grounded one in the RF region and acts as a microwave ring trap for electrons. We set the widths of the GND and RF electrodes to be $160$~$\mu$m and $80$~$\mu$m, respectively, and that of the MW electrodes at the cross-section to be 30~$\mu$m.

As for an ion, on one hand, the trapping potential is nothing but a linear RF Paul trap, since the ion cannot follow the microwave field.  On the other hand, an electron is trapped by a ring-like microwave Paul trap. 
Normalized pseudo-potential profiles for an ion and an electron in the cross-section along the black dotted line in Fig.~\ref{Fig5}(a) are respectively shown in Figs.~\ref{Fig5}(b) and (c). The trapping potential for an electron becomes 40~meV-depth and yields the secular frequency of 800~MHz with an applied 4~GHz-microwave voltage of 20~V. 
The trapping potential for an ion is estimated to be 20~meV-depth with secular frequency of 3~MHz by assuming the use of a beryllium ion and an application of 40~MHz-RF voltage of 30~V, which is sufficient for ion-trap experiments~\cite{Leibrandt2009-ij}. \textcolor{black}{Here the grounded ceiling is located at 200 $\mu$m height.}

Due to the slight difference of the electrode dimension, electrons and ions are trapped at different heights. The electrode design shown in Fig.~\ref{Fig5}(b,c) yields the difference of their heights by $50$~$\mu$m and this difference can be made smaller down to $10$~$\mu$m if we add a grounded ceiling plate at 100~$\mu$m above the trap to modify the pseudo-potential profiles, see Figs.~\ref{Fig5}(d) and (e).  This situation, where the electrons and ions are trapped close but apart, is beneficial for preventing them from recombination while keeping sizable Coulomb interaction between them. The coupling between an electron and an ion simultaneously trapped with our proposed setup will be discussed in the following Sections.

\subsection{Readout of motional state using trapped ions}

As has been proposed in the previous Section, an electron and an ion can be trapped by applied ac voltages separately but close to each other.  Here let us consider the interaction between them, starting from total potential energy $U$ of the electron-ion hybrid system
\begin{align}
    U=\frac{1}{2}m_\mathrm{e}\omega_\mathrm{e}^2 x^2+\frac{1}{2}m_\mathrm{i}\omega_\mathrm{i}^2 y^2 + V,
\end{align} 
including the Coulomb energy
\begin{align}
    V = - \frac{e^2}{4\pi\epsilon_0}\frac{1}{L-y+x}.
\end{align}
In above expressions, an electron and an ion is trapped at positions separated by $L$ and displacements from their trapping points are denoted as $x$ for the electron and $y$ for the ion. With subscripts $\mathrm{e}$ for the electron and $\mathrm{i}$ for the ion attached, $m$ and $\omega$ represent the mass and secular frequency of each particle, respectively.  Well-known physical constants of $e$, the elementary charge, and $\epsilon_0$, the vacuum permittivity, are also used. 

\begin{table}[t]
\begin{tabular}{cccccc}
\hline
$\omega_\mathrm{e}/2\pi$ (MHz)& $L$ ($\mathrm{\mu m}$)& $g_0/2\pi$ (kHz) & $\alpha/2\pi$ (kHz)  & $\bar{n}_\mathrm{e}$\\
\hline \hline
 800 & 10 &  33& $-33$ &  $5.3 \times 10^{-2}$ \\
 800 & $50$ & $0.39$& $-34$ &  5.9 \\
 500 & $10$ & $39$ & $-2.6\times10^3$ &  $3.8 \times 10^{-2}$ \\
 500 & 7 &  $1.6\times10^3$& $-2.5\times10^3$ &  $2.2 \times 10^{-3}$ \\
 \hline
\end{tabular}
\caption{Coupling strengths of the electron-ion hybrid system for various parameters.  $g_0$: coupling strength of the optomechanical term, $\alpha$: strength of the self-Kerr interaction, $\bar{n}_\mathrm{e}$: estimated phonon number of a trapped electron. $\omega_\mathrm{i}/2\pi = 2$~MHz is assumed in common. }\label{table1}
\end{table}

Coulomb attraction $V$ makes the electron and the ion bring themselves closer and the equilibrium points are shifted from the original trapping point.  Moreover, the potential is distorted to get anharmonic since the Coulomb interaction is nonlinear by nature. We shall expand $V$ under the assumption that the relative displacement is much smaller than the electron-ion distance, $\abs{x-y} \ll L$. This expansion yields polynomials of $x$ and $y$. The terms proportional to $x$ and $y$ result in the shift of positions and those to $x^2$ and $y^2$ lead to the shift of oscillation frequencies.  The term $xy$ contains a beam-splitter or two-mode-squeezing interactions. Higher-order terms $x^3$, $x^2y$, $xy^2$ and $y^3$ stand for the second-order nonlinearities and $x^4$, $y^4$, $\dots$ represent the third-order nonlinearities. Since we consider the situation that the phonon frequencies of the electron and the ion are different, the beam-splitter and two-mode-squeezing interactions are not valid here.  Likewise the rapidly oscillating terms are omitted and the series expansion is truncated in our analysis to get a Hamiltonian of the whole system as
\begin{align}
    \hat{H} = \hbar \omega_\mathrm{e} \hat{a}^\dagger \hat{a} + \hbar \omega_\mathrm{i} \hat{c}^\dagger \hat{c} - \hbar g_0 \hat{a}^\dagger \hat{a} (\hat{c}^\dagger + \hat{c}) - \frac{\hbar \alpha}{2} \hat{a}^\dagger \hat{a}^\dagger \hat{a} \hat{a}, \label{H}
\end{align}
where $\omega_\mathrm{e}$ now stands for the renormalized secular frequency of a trapped electron and $\omega_\mathrm{i}$ is the one for the trapped ion.  The ladder operators of the phonon mode of the trapped ion are denoted by $\hat{c}$ and $\hat{c}^\dagger$. 
The third term is much like an cavity optomechanical interaction, where the electron mimics the cavity and the ion plays the role of a mechanical oscillator.  The coupling strength $g_0$ is given as\textcolor{black}{
\begin{align}
    g_0 = \frac{1}{\hbar} \left( \frac{e^2}{4\pi \epsilon_0} \frac{6 x_\mathrm{zpf}^2 y_\mathrm{zpf}}{L^4} - \frac{2\hbar g_{\textcolor{black}{\mathrm{C}}} \beta}{\omega_\mathrm{e}-\omega_\mathrm{i}} \right)
\end{align}
}
using the zero-point fluctuations $x_\mathrm{zpf} = \sqrt{\hbar/2m_\mathrm{e} \omega_\mathrm{e}}$ and $y_\mathrm{zpf} = \sqrt{\hbar/2m_\mathrm{i} \omega_\mathrm{i}}$. 
\textcolor{black}{In addition to the nonlinearity arising from the Coulomb interaction, we included the nonlinearity of electron motion arising from the effective potential itself in the second term. Along with this correction, we introduced $g_{\textcolor{black}{\mathrm{C}}} = (1/\hbar)(\partial^2 V/ \partial x \partial y)x_\mathrm{zpf}y_\mathrm{zpf}$. $\beta$ is extracted from the second-order nonlinearity of the effective potential for the electron.}
Though the interaction term is principally given by the optomechanical one, we include the fourth, self-Kerr term for it limits the driven number of phonons, as described in later paragraph. 
\textcolor{black}{The constant $\alpha$ reads
\begin{align}
    \alpha = \alpha_{\textcolor{black}{\mathrm{C}}} + \alpha_{\textcolor{black}{\mathrm{K}}} - \frac{6\beta^2}{\omega_\mathrm{e}}
\end{align}
where
\begin{align}
    \alpha_\mathrm{C} &= \frac{1}{\hbar} \frac{e^2}{4\pi \epsilon_0} \frac{12 x_\mathrm{zpf}^4 }{L^5}
\end{align}
is the third-order nonlinearity and $\alpha_\mathrm{K}$ the one from the effective potential for the electron. The Hamiltonian in Eq.~(\ref{H}) is obtained in the same manner as was done in Ref.~\cite{Noguchi2020-sd}.}

Just as in the cavity optomechanics~\cite{Aspelmeyer2014-ew}, the interaction term $\hbar g_0 \hat{a}^\dagger \hat{a} (\hat{c}^\dagger + \hat{c})$ allows for the three-wave mixing, one of which is the phonon of an ion and two are the phonon of an electron.  If we drive the electron motion with microwaves of its frequency being $\omega_\mathrm{e} + \omega_\mathrm{i}$, the linearized interaction appears as $\hbar g_0 \sqrt{n_\mathrm{d}} (\hat{a}^\dagger \hat{c}^\dagger +  \hat{a} \hat{c} )$, with $n_\mathrm{d}$ being the mean number of phonons of the electron. Driving field is assumed to be intense so that its amplitude can be thought of the mean phonon number of the driven mode.   This two-mode-squeezing interaction
amplifies the motion of the electron, whose state can be readout through a thermometry of the trapped ion.

The motion of the electron is driven in such a scheme, however, the self-Kerr term $(\hbar \alpha/2) \hat{a}^\dagger \hat{a}^\dagger \hat{a} \hat{a}$ imposes small but finite energy shift of the number states with large occupation numbers.  This limits the mean number of phonons that can be excited by the drive tone, or in other words the effect of the driving saturates. The saturation occurs roughly when the frequency shift $\alpha n_\mathrm{d}$ gets comparable to the driving strength itself, $g = g_0 \sqrt{n_\mathrm{d}}$.  This tells us the limit of coupling strength $g_\mathrm{max}$ of $g$ as
\begin{align}
    g_\mathrm{max} = \frac{g_0^2}{\alpha}.
\end{align}
For various trapping parameters, we calculate the coupling strength $g_0$ and nonlinear coefficient $\alpha$ in Table~\ref{table1}. In all these parameters, the driven oscillatory motion exhibits the oscillation amplitude of no more than a few micrometers, \textcolor{black}{however, driven number of phonons is no more than 1 since $g$ and $\alpha$ are comparable in their magnitudes}. The motional state of the ion can be identified using its optical transition and even that of the electron is possible with this scheme.

\subsection{Sympathetic cooling with trapped ion}

When we drive the electron with $\omega_\mathrm{e} - \omega_\mathrm{i}$,
the interaction term becomes \textcolor{black}{$\hbar g (\hat{a}^{\dagger} \hat{c} +  \hat{a} \hat{c}^{\dagger} )$}.  Suppose that the cooling rate of the ion's motion is $\Gamma_\mathrm{i}$.  Through the beam-splitter interaction synthesized above, the electron is sympathetically cooled~\cite{Itano_1995} by a cooling rate of \textcolor{black}{$\Gamma_\downarrow^{\mathrm{e}} = 4 g^{2}/\Gamma_\mathrm{i}$}, where \textcolor{black}{$\Gamma_\mathrm{i} > g$} is assumed.  Then the average phonon number of the electron is estimated by $\bar{n}_\mathrm{e} = n_\mathrm{th} {\Gamma_\mathrm{th}^{\mathrm{e}}}^{\prime}/(\Gamma_\downarrow^{\mathrm{e}} + {\Gamma_\mathrm{th}^{\mathrm{e}}}^{\prime})$, where $n_\mathrm{th} = 12$ is the mean phonon number of the electron at 300~mK with the phonon frequency of 500~MHz and ${\Gamma_\mathrm{th}^{\mathrm{e}}}^{\prime} = \Gamma_\mathrm{th}^{\mathrm{e}}+\Gamma_\mathrm{i}$ with $\Gamma_\mathrm{th}^{\mathrm{e}}$ being the coupling rate of electron's motion to the thermal bath, which we shall conservatively assume it to be 10~Hz~\cite{Yu2022-zz}. 

The values of $g_\mathrm{max}$ and the final occupation number $\bar{n}_\mathrm{e}$ are listed in Table.~\ref{table1} for various parameters of the electron-ion coupled system with $\Gamma_\mathrm{i}/2\pi = 10$~kHz being assumed as a typical value.  The final occupation number $\bar{n}_\mathrm{e}$ can be well below unity by the thermalization with the ion in most conditions listed in the Table, which confirms the validity of sympathetic cooling of a trapped electron down to the motional ground state using a trapped ion. 
One concern is the possible heating of electrons (ions) due to the presence of the RF for trapping ions (electrons) in the two-frequency Paul trap, which should be investigated with the actual implementation of it in future work.
One more thing that we would like to mention is that this method, the parametrically-driven sympathetic cooling of mass-imbalanced charged particles, is useful not only for the cooling of electrons but also in the ion-trap experiments, since our method allow us to avoid inefficiency of the sympathetic cooling of mass-imbalanced ions by bringing their motional frequency effectively on resonance.

\section{Conclusion} \label{conclusion}

As a summary, we proposed and analyzed hybrid quantum systems consisting of a trapped electron interacting with superconducting circuits and a trapped ion. The basic idea of the electron trap using multiple modes of the $\lambda/4$ resonator and the cryocooler-compatible, low-energy electron source are introduced and proved to be valid for the use in proposed electron-trapping experiments.  We further revealed that the light mass of the electron and hence the high secular-motion frequency of the trapped one result in the strong electron-circuit and electron-ion couplings, where the latter is aided by the nonlinearity of the Coulomb interaction. In both hybrid quantum systems, the single-phonon-level readout and the ground-state cooling of the motional state of the trapped electron are feasible by microwave and optical means.  Combined with the results studied in Ref.~\cite{Yu2022-zz}, the highly-efficient and high-fidelity quantum operations are available in the trapped-electron system, and this novel system manifests itself as a new playground for the development of quantum technologies.

We acknowledge Genya Watanabe, Shotaro Shirai and Yasunobu Nakamura for fruitful discussions. KT acknowledges SPRING GX and Q-step programs, and MS does WINGS-ABC program. This work is supported by JST ERATO MQM project (Grant No. JPMJER1601), JSPS KAKENHI (Grant No. 19H01821) and JST SPRING (Grant No. JPMJSP2108).

\bibliography{apssamp}

@PREAMBLE{
 "\providecommand{\noopsort}[1]{}" 
 # "\providecommand{\singleletter}[1]{#1}
}


@ARTICLE{Foot2018-gl,
  author   = "Foot, C J and Trypogeorgos, D and Bentine, E and Gardner, A and
              Keller, M",
  abstract = "We describe the operation of an electrodynamic ion trap in which
              the electric quadrupole field oscillates at two frequencies. This
              mode of operation allows simultaneous tight confinement of ions
              with extremely different charge-to-mass ratios, e.g., singly
              ionised atomic ions together with multiply charged nanoparticles.
              We derive the stability conditions for two-frequency operation
              from asymptotic properties of the solutions of the Mathieu
              equation and give a general treatment of the effect of damping on
              parametric resonances. Two-frequency operation is effective when
              the two species' mass ratios and charge ratios are sufficiently
              large, and further when the frequencies required to optimally
              trap each species are widely separated. This system resembles two
              coincident Paul traps, each operating close to a frequency
              optimised for one of the species, such that both species are
              tightly confined. This method of operation provides an advantage
              over single-frequency Paul traps, in which the more weakly
              confined species forms a sheath around a central core of tightly
              confined ions. We verify these ideas using numerical simulations
              and by measuring the parametric heating induced in experiments by
              the additional driving frequency.",
  journal  = "Int. J. Mass Spectrom.",
  volume   =  430,
  pages    = "117--125",
  month    =  jul,
  year     =  2018,
  keywords = "Ion trap; Multiple frequencies; Parametric heating"
}

@ARTICLE{Dehmelt1995-fs,
  author    = "Dehmelt, Hans",
  abstract  = "Estimates of operating parameters of a Paul trap suitable for
               synthesis and spectroscopy of anti-molecular hydrogen ions are
               presented. The trap may employ a 2-frequency trapping voltage.
               Transportable cryogenic Paul traps for the long-time storage of
               positrons and antiprotons may make such experiments possible in
               an average university lab in the not too distant future.",
  journal   = "Phys. Scr.",
  publisher = "IOP Publishing",
  volume    =  1995,
  number    = "T59",
  pages     = "423",
  month     =  jan,
  year      =  1995,
  language  = "en"
}

@ARTICLE{Leefer2016-du,
  author   = "Leefer, Nathan and Krimmel, Kai and Bertsche, William and Budker,
              Dmitry and Fajans, Joel and Folman, Ron and H{\"a}ffner, Hartmut
              and Schmidt-Kaler, Ferdinand",
  abstract = "Radio-frequency (rf) Paul traps operated with multifrequency rf
              trapping potentials provide the ability to independently confine
              charged particle species with widely different charge-to-mass
              ratios. In particular, these traps may find use in the field of
              antihydrogen recombination, allowing antiproton and positron
              clouds to be trapped and confined in the same volume without the
              use of large superconducting magnets. We explore the stability
              regions of two-frequency Paul traps and perform numerical
              simulations of small samples of multispecies charged-particle
              mixtures of up to twelve particles that indicate the promise of
              these traps for antihydrogen recombination.",
  journal  = "Hyperfine Interact.",
  volume   =  238,
  number   =  1,
  pages    = "12",
  month    =  dec,
  year     =  2016
}

@ARTICLE{Goryachev2014-ep,
  author    = "Goryachev, Maxim and Farr, Warrick G and Creedon, Daniel L and Fan, Yaohui and Kostylev, Mikhail and Tobar, Michael E",
  journal   = "Phys. Rev. Applied",
  publisher = "American Physical Society",
  volume    =  2,
  number    =  5,
  pages     = "054002",
  month     =  nov,
  year      =  2014
}

@ARTICLE{Molmer1999-dq,
  author    = "M{\o}lmer, Klaus and S{\o}rensen, Anders",
  journal   = "Phys. Rev. Lett.",
  publisher = "American Physical Society",
  volume    =  82,
  number    =  9,
  pages     = "1835--1838",
  month     =  mar,
  year      =  1999
}

@ARTICLE{Zwerver2022-xl,
  author    = "Zwerver, A M J and Kr{\"a}henmann, T and Watson, T F and
               Lampert, L and George, H C and Pillarisetty, R and Bojarski, S A
               and Amin, P and Amitonov, S V and Boter, J M and Caudillo, R and
               Corras-Serrano, D and Dehollain, J P and Droulers, G and Henry,
               E M and Kotlyar, R and Lodari, M and L{\"u}thi, F and Michalak,
               D J and Mueller, B K and Neyens, S and Roberts, J and
               Samkharadze, N and Zheng, G and Zietz, O K and Scappucci, G and
               Veldhorst, M and Vandersypen, L M K and Clarke, J S",
  abstract  = "Full-scale quantum computers require the integration of millions
               of qubits, and the potential of using industrial semiconductor
               manufacturing to meet this need has driven the development of
               quantum computing in silicon quantum dots. However, fabrication
               has so far relied on electron-beam lithography and, with a few
               exceptions, conventional lift-off processes that suffer from low
               yield and poor uniformity. Here we report quantum dots that are
               hosted at a 28Si/28SiO2 interface and fabricated in a 300 mm
               semiconductor manufacturing facility using all-optical
               lithography and fully industrial processing. With this approach,
               we achieve nanoscale gate patterns with excellent yield. In the
               multi-electron regime, the quantum dots allow good tunnel
               barrier control---a crucial feature for fault-tolerant two-qubit
               gates. Single-spin qubit operation using magnetic resonance in
               the few-electron regime reveals relaxation times of over 1 s at
               1 T and coherence times of over 3 ms. Silicon spin qubits can be
               fabricated in a 300 mm semiconductor manufacturing facility
               using all-optical lithography and fully industrial processing.",
  journal   = "Nature Electronics",
  publisher = "Nature Publishing Group",
  volume    =  5,
  number    =  3,
  pages     = "184--190",
  month     =  mar,
  year      =  2022,
  language  = "en"
}

@ARTICLE{Foxen2020-ti,
  author   = "Foxen, B and Neill, C and Dunsworth, A and Roushan, P and Chiaro,
              B and Megrant, A and Kelly, J and Chen, Zijun and Satzinger, K
              and Barends, R and Arute, F and Arya, K and Babbush, R and Bacon,
              D and Bardin, J C and Boixo, S and Buell, D and Burkett, B and
              Chen, Yu and Collins, R and Farhi, E and Fowler, A and Gidney, C
              and Giustina, M and Graff, R and Harrigan, M and Huang, T and
              Isakov, S V and Jeffrey, E and Jiang, Z and Kafri, D and
              Kechedzhi, K and Klimov, P and Korotkov, A and Kostritsa, F and
              Landhuis, D and Lucero, E and McClean, J and McEwen, M and Mi, X
              and Mohseni, M and Mutus, J Y and Naaman, O and Neeley, M and
              Niu, M and Petukhov, A and Quintana, C and Rubin, N and Sank, D
              and Smelyanskiy, V and Vainsencher, A and White, T C and Yao, Z
              and Yeh, P and Zalcman, A and Neven, H and Martinis, J M and
              {Google AI Quantum}",
  abstract = "Quantum algorithms offer a dramatic speedup for computational
              problems in material science and chemistry. However, any
              near-term realizations of these algorithms will need to be
              optimized to fit within the finite resources offered by existing
              noisy hardware. Here, taking advantage of the adjustable coupling
              of gmon qubits, we demonstrate a continuous two-qubit gate set
              that can provide a threefold reduction in circuit depth as
              compared to a standard decomposition. We implement two gate
              families: an imaginary swap-like (iSWAP-like) gate to attain an
              arbitrary swap angle, $\vartheta$, and a controlled-phase gate
              that generates an arbitrary conditional phase, ϕ. Using one of
              each of these gates, we can perform an arbitrary two-qubit gate
              within the excitation-preserving subspace allowing for a complete
              implementation of the so-called Fermionic simulation (fSim) gate
              set. We benchmark the fidelity of the iSWAP-like and
              controlled-phase gate families as well as 525 other fSim gates
              spread evenly across the entire fSim($\vartheta$,ϕ) parameter
              space, achieving a purity-limited average two-qubit Pauli error
              of 3.8$\times$10^\{-3\} per fSim gate.",
  journal  = "Phys. Rev. Lett.",
  volume   =  125,
  number   =  12,
  pages    = "120504",
  month    =  sep,
  year     =  2020,
  language = "en"
}

@ARTICLE{Gaebler2016-rd,
  author   = "Gaebler, J P and Tan, T R and Lin, Y and Wan, Y and Bowler, R and
              Keith, A C and Glancy, S and Coakley, K and Knill, E and
              Leibfried, D and Wineland, D J",
  journal  = "Phys. Rev. Lett.",
  volume   =  117,
  number   =  6,
  pages    = "060505",
  month    =  aug,
  year     =  2016,
  language = "en"
}

@ARTICLE{Aspelmeyer2014-ew,
  author    = "Aspelmeyer, Markus and Kippenberg, Tobias J and Marquardt,
               Florian",
  journal   = "Rev. Mod. Phys.",
  publisher = "American Physical Society",
  volume    =  86,
  number    =  4,
  pages     = "1391--1452",
  month     =  dec,
  year      =  2014
}

@ARTICLE{Wallraff2004-fa,
  author   = "Wallraff, A and Schuster, D I and Blais, A and Frunzio, L and
              Huang, R- S and Majer, J and Kumar, S and Girvin, S M and
              Schoelkopf, R J",
  abstract = "The interaction of matter and light is one of the fundamental
              processes occurring in nature, and its most elementary form is
              realized when a single atom interacts with a single photon.
              Reaching this regime has been a major focus of research in atomic
              physics and quantum optics for several decades and has generated
              the field of cavity quantum electrodynamics. Here we perform an
              experiment in which a superconducting two-level system, playing
              the role of an artificial atom, is coupled to an on-chip cavity
              consisting of a superconducting transmission line resonator. We
              show that the strong coupling regime can be attained in a
              solid-state system, and we experimentally observe the coherent
              interaction of a superconducting two-level system with a single
              microwave photon. The concept of circuit quantum electrodynamics
              opens many new possibilities for studying the strong interaction
              of light and matter. This system can also be exploited for
              quantum information processing and quantum communication and may
              lead to new approaches for single photon generation and
              detection.",
  journal  = "Nature",
  volume   =  431,
  number   =  7005,
  pages    = "162--167",
  month    =  sep,
  year     =  2004,
  language = "en"
}

@ARTICLE{Kimble2008-bc,
  author   = "Kimble, H J",
  abstract = "Quantum networks provide opportunities and challenges across a
              range of intellectual and technical frontiers, including quantum
              computation, communication and metrology. The realization of
              quantum networks composed of many nodes and channels requires new
              scientific capabilities for generating and characterizing quantum
              coherence and entanglement. Fundamental to this endeavour are
              quantum interconnects, which convert quantum states from one
              physical system to those of another in a reversible manner. Such
              quantum connectivity in networks can be achieved by the optical
              interactions of single photons and atoms, allowing the
              distribution of entanglement across the network and the
              teleportation of quantum states between nodes.",
  journal  = "Nature",
  volume   =  453,
  number   =  7198,
  pages    = "1023--1030",
  month    =  jun,
  year     =  2008,
  language = "en"
}

@ARTICLE{Kitaev2003-so,
  author   = "Kitaev, A Yu",
  abstract = "A two-dimensional quantum system with anyonic excitations can be
              considered as a quantum computer. Unitary transformations can be
              performed by moving the excitations around each other.
              Measurements can be performed by joining excitations in pairs and
              observing the result of fusion. Such computation is
              fault-tolerant by its physical nature.",
  journal  = "Ann. Phys.",
  volume   =  303,
  number   =  1,
  pages    = "2--30",
  month    =  jan,
  year     =  2003
}

@ARTICLE{Bravyi1998-jl,
  author        = "Bravyi, S B and Yu. Kitaev, A",
  journal  = "arXiv:quant-ph/9811052v1",
}

@article{Weber,
author = {J. R. Weber  and W. F. Koehl  and J. B. Varley  and A. Janotti  and B. B. Buckley  and C. G. Van de Walle  and D. D. Awschalom },
journal = {Proceedings of the National Academy of Sciences},
volume = {107},
number = {19},
pages = {8513-8518},
year = {2010},
abstract = {Identifying and designing physical systems for use as qubits, the basic units of quantum information, are critical steps in the development of a quantum computer. Among the possibilities in the solid state, a defect in diamond known as the nitrogen-vacancy (NV-1) center stands out for its robustness—its quantum state can be initialized, manipulated, and measured with high fidelity at room temperature. Here we describe how to systematically identify other deep center defects with similar quantum-mechanical properties. We present a list of physical criteria that these centers and their hosts should meet and explain how these requirements can be used in conjunction with electronic structure theory to intelligently sort through candidate defect systems. To illustrate these points in detail, we compare electronic structure calculations of the NV-1 center in diamond with those of several deep centers in 4H silicon carbide (SiC). We then discuss the proposed criteria for similar defects in other tetrahedrally coordinated semiconductors.}}

@ARTICLE{Kjaergaard2020-er,
  author    = "Kjaergaard, Morten and Schwartz, Mollie E and Braum{\"u}ller,
               Jochen and Krantz, Philip and Wang, Joel I-J and Gustavsson,
               Simon and Oliver, William D",
  abstract  = "Superconducting qubits are leading candidates in the race to
               build a quantum computer capable of realizing computations
               beyond the reach of modern supercomputers. The superconducting
               qubit modality has been used to demonstrate prototype algorithms
               in the noisy intermediate-scale quantum (NISQ) technology era,
               in which non-error-corrected qubits are used to implement
               quantum simulations and quantum algorithms. With the recent
               demonstrations of multiple high-fidelity, two-qubit gates as
               well as operations on logical qubits in extensible
               superconducting qubit systems, this modality also holds promise
               for the longer-term goal of building larger-scale
               error-corrected quantum computers. In this brief review, we
               discuss several of the recent experimental advances in qubit
               hardware, gate implementations, readout capabilities, early NISQ
               algorithm implementations, and quantum error correction using
               superconducting qubits. Although continued work on many aspects
               of this technology is certainly necessary, the pace of both
               conceptual and technical progress in recent years has been
               impressive, and here we hope to convey the excitement stemming
               from this progress.",
  journal   = "Annu. Rev. Condens. Matter Phys.",
  publisher = "Annual Reviews",
  volume    =  11,
  number    =  1,
  pages     = "369--395",
  month     =  mar,
  year      =  2020
}

@ARTICLE{Saffman2016-sn,
  author    = "Saffman, M",
  abstract  = "We present a review of quantum computation with neutral atom
               qubits. After an overview of architectural options and
               approaches to preparing large qubit arrays we examine Rydberg
               mediated gate protocols and fidelity for two- and multi-qubit
               interactions. Quantum simulation and Rydberg dressing are
               alternatives to circuit based quantum computing for exploring
               many body quantum dynamics. We review the properties of the
               dressing interaction and provide a quantitative figure of merit
               for the complexity of the coherent dynamics that can be accessed
               with dressing. We conclude with a summary of the current status
               and an outlook for future progress.",
  journal   = "J. Phys. B At. Mol. Opt. Phys.",
  publisher = "IOP Publishing",
  volume    =  49,
  number    =  20,
  pages     = "202001",
  month     =  oct,
  year      =  2016,
  language  = "en"
}

@ARTICLE{Gulde2001-dc,
  author    = "Gulde, S and Rotter, D and Barton, P and Schmidt-Kaler, F and
               Blatt, R and Hogervorst, W",
  journal   = "Appl. Phys. B",
  publisher = "Springer Science and Business Media LLC",
  volume    =  73,
  number    =  8,
  pages     = "861--863",
  month     =  dec,
  year      =  2001,
  language  = "en"
}

@ARTICLE{Matthiesen2021-mf,
  author    = "Matthiesen, Clemens and Yu, Qian and Guo, Jinen and Alonso,
               Alberto M and H{\"a}ffner, Hartmut",
  journal   = "Phys. Rev. X",
  publisher = "American Physical Society",
  volume    =  11,
  number    =  1,
  pages     = "011019",
  month     =  jan,
  year      =  2021
}

@ARTICLE{Gely2019-mz,
  author   = "Gely, Mario F and Kounalakis, Marios and Dickel, Christian and
              Dalle, Jacob and Vatr{\'e}, R{\'e}my and Baker, Brian and
              Jenkins, Mark D and Steele, Gary A",
  abstract = "Detecting weak radio-frequency electromagnetic fields plays a
              crucial role in a wide range of fields, from radio astronomy to
              nuclear magnetic resonance imaging. In quantum optics, the
              ultimate limit of a weak field is a single photon. Detecting and
              manipulating single photons at megahertz frequencies presents a
              challenge because, even at cryogenic temperatures, thermal
              fluctuations are appreciable. Using a gigahertz superconducting
              qubit, we observed the quantization of a megahertz
              radio-frequency resonator, cooled it to the ground state, and
              stabilized Fock states. Releasing the resonator from our control,
              we observed its rethermalization with nanosecond resolution.
              Extending circuit quantum electrodynamics to the megahertz
              regime, we have enabled the exploration of thermodynamics at the
              quantum scale and allowed interfacing quantum circuits with
              megahertz systems such as spin systems or macroscopic mechanical
              oscillators.",
  journal  = "Science",
  volume   =  363,
  number   =  6431,
  pages    = "1072--1075",
  month    =  mar,
  year     =  2019,
  language = "en"
}

@ARTICLE{Yu2022-zz,
  author    = "Yu, Qian and Alonso, Alberto M and Caminiti, Jackie and Beck,
               Kristin M and Sutherland, R Tyler and Leibfried, Dietrich and
               Rodriguez, Kayla J and Dhital, Madhav and Hemmerling, Boerge and
               H{\"a}ffner, Hartmut",
  journal   = "Phys. Rev. A",
  publisher = "American Physical Society",
  volume    =  105,
  number    =  2,
  pages     = "022420",
  month     =  feb,
  year      =  2022
}

@ARTICLE{Ruster2016-jp,
  author   = "Ruster, T and Schmiegelow, C T and Kaufmann, H and Warschburger,
              C and Schmidt-Kaler, F and Poschinger, U G",
  abstract = "We demonstrate unprecedentedly long lifetimes for electron spin
              superposition states of a single trapped $$^\{40\}$$Ca$$^+$$ion.
              For a Ramsey measurement, we achieve a
              $$1\{/\}\textbackslashsqrt\{e\}$$coherence time of 300(50) ms,
              while a spin-echo experiment yields a coherence time of 2.1(1) s.
              The latter corresponds to residual effective rms magnetic field
              fluctuations $$\{\textbackslashle \}2.7\textbackslashtimes
              10^\{-12\}\textbackslash,\textbackslashhbox \{T\}$$during a
              measurement time of about 1500 s. The suppression of decoherence
              induced by fluctuating magnetic fields is achieved by combining a
              two-layer $$\textbackslashmu$$-metal shield, which reduces
              external magnetic noise by 20--30 dB for frequencies of 50
              Hz--100 kHz, with Sm$$_2$$Co$$_\{17\}$$permanent magnets for
              generating a quantizing magnetic field of 0.37 mT. Our results
              extend the coherence time of the simple-to-operate trapped-ion
              spin qubit to ultralong coherence times which so far have been
              observed only for magnetic insensitive transitions in atomic
              qubits with hyperfine structure.",
  journal  = "Appl. Phys. B",
  volume   =  122,
  number   =  10,
  pages    = "254",
  month    =  sep,
  year     =  2016
}

@ARTICLE{Bruzewicz2019-ac,
  author    = "Bruzewicz, Colin D and Chiaverini, John and McConnell, Robert
               and Sage, Jeremy M",
  abstract  = "Trapped ions are among the most promising systems for practical
               quantum computing (QC). The basic requirements for universal QC
               have all been demonstrated with ions, and quantum algorithms
               using few-ion-qubit systems have been implemented. We review the
               state of the field, covering the basics of how trapped ions are
               used for QC and their strengths and limitations as qubits. In
               addition, we discuss what is being done, and what may be
               required, to increase the scale of trapped ion quantum computers
               while mitigating decoherence and control errors. Finally, we
               explore the outlook for trapped-ion QC. In particular, we
               discuss near-term applications, considerations impacting the
               design of future systems of trapped ions, and experiments and
               demonstrations that may further inform these considerations.",
  journal   = "Applied Physics Reviews",
  publisher = "American Institute of Physics",
  volume    =  6,
  number    =  2,
  pages     = "021314",
  month     =  jun,
  year      =  2019
}

@ARTICLE{Brewer2019-jh,
  author   = "Brewer, S M and Chen, J-S and Hankin, A M and Clements, E R and
              Chou, C W and Wineland, D J and Hume, D B and Leibrandt, D R",
  journal  = "Phys. Rev. Lett.",
  volume   =  123,
  number   =  3,
  pages    = "033201",
  month    =  jul,
  year     =  2019,
  language = "en"
}

@ARTICLE{Wineland1973-se,
  author    = "Wineland, D and Ekstrom, P and Dehmelt, H",
  journal   = "Phys. Rev. Lett.",
  publisher = "American Physical Society (APS)",
  volume    =  31,
  number    =  21,
  pages     = "1279--1282",
  month     =  nov,
  year      =  1973,
  copyright = "http://link.aps.org/licenses/aps-default-license"
}

@article{Reinhold_2020,
	doi = {10.1038/s41567-020-0931-8},
  
	url = {https://doi.org/10.1038
  
	year = 2020,
	month = {jun},
  
	publisher = {Springer Science and Business Media {LLC}
},
  
	volume = {16},
  
	number = {8},
  
	pages = {822--826},
  
	author = {Philip Reinhold and Serge Rosenblum and Wen-Long Ma and Luigi Frunzio and Liang Jiang and Robert J. Schoelkopf},
  
	journal = {Nature Physics}
}

@ARTICLE{Schuster2007-ii,
  author   = "Schuster, D I and Houck, A A and Schreier, J A and Wallraff, A
              and Gambetta, J M and Blais, A and Frunzio, L and Majer, J and
              Johnson, B and Devoret, M H and Girvin, S M and Schoelkopf, R J",
  abstract = "Electromagnetic signals are always composed of photons, although
              in the circuit domain those signals are carried as voltages and
              currents on wires, and the discreteness of the photon's energy is
              usually not evident. However, by coupling a superconducting
              quantum bit (qubit) to signals on a microwave transmission line,
              it is possible to construct an integrated circuit in which the
              presence or absence of even a single photon can have a dramatic
              effect. Such a system can be described by circuit quantum
              electrodynamics (QED)-the circuit equivalent of cavity QED, where
              photons interact with atoms or quantum dots. Previously, circuit
              QED devices were shown to reach the resonant strong coupling
              regime, where a single qubit could absorb and re-emit a single
              photon many times. Here we report a circuit QED experiment in the
              strong dispersive limit, a new regime where a single photon has a
              large effect on the qubit without ever being absorbed. The
              hallmark of this strong dispersive regime is that the qubit
              transition energy can be resolved into a separate spectral line
              for each photon number state of the microwave field. The strength
              of each line is a measure of the probability of finding the
              corresponding photon number in the cavity. This effect is used to
              distinguish between coherent and thermal fields, and could be
              used to create a photon statistics analyser. As no photons are
              absorbed by this process, it should be possible to generate
              non-classical states of light by measurement and perform
              qubit-photon conditional logic, the basis of a logic bus for a
              quantum computer.",
  journal  = "Nature",
  volume   =  445,
  number   =  7127,
  pages    = "515--518",
  month    =  feb,
  year     =  2007,
  language = "en"
}

@article{Daniilidis_2013,
	doi = {10.1088/1367-2630/15/7/073017},
	url = {https://doi.org/10.1088/1367-2630/15/7/073017},
	year = 2013,
	month = {jul},
	publisher = {{IOP} Publishing},
	volume = {15},
	number = {7},
	pages = {073017},
	author = {Nikos Daniilidis and Dylan J Gorman and Lin Tian and Hartmut Häffner},
	title = {Quantum information processing with trapped electrons and superconducting electronics},
	journal = {New Journal of Physics},
	abstract = {We describe a parametric frequency conversion scheme for trapped charged particles, which enables a coherent interface between atomic and solid-state quantum systems. The scheme uses geometric nonlinearities of the potential of coupling electrodes near a trapped particle, and can be implemented using standard charged-particle traps. Our scheme does not rely on actively driven solid-state devices, and is hence largely immune to noise in such devices. We present a toolbox which can be used to build electron-based quantum information processing platforms, as well as quantum hybrid platforms using trapped electrons and superconducting electronics.}
}

@article{Kotler_2017,
  title = {Hybrid quantum systems with trapped charged particles},
  author = {Kotler, Shlomi and Simmonds, Raymond W. and Leibfried, Dietrich and Wineland, David J.},
  journal = {Phys. Rev. A},
  volume = {95},
  issue = {2},
  pages = {022327},
  numpages = {29},
  year = {2017},
  month = {Feb},
  publisher = {American Physical Society},
  doi = {10.1103/PhysRevA.95.022327},
  url = {https://link.aps.org/doi/10.1103/PhysRevA.95.022327}
}

@article{Peng_2017,
  author = {Peng, Pai and Matthiesen, Clemens and H\"affner, Hartmut},
  journal = {Phys. Rev. A},
  volume = {95},
  issue = {1},
  pages = {012312},
  numpages = {8},
  year = {2017},
  month = {Jan},
  publisher = {American Physical Society},
  doi = {10.1103/PhysRevA.95.012312},
  url = {https://link.aps.org/doi/10.1103/PhysRevA.95.012312}
}

@article{Wineland_1975,
author = {Wineland,D. J.  and Dehmelt,H. G. },
title = {Principles of the stored ion calorimeter},
journal = {Journal of Applied Physics},
volume = {46},
number = {2},
pages = {919-930},
year = {1975},
doi = {10.1063/1.321602},

URL = { 
        https://doi.org/10.1063/1.321602
    
},
eprint = { 
        https://doi.org/10.1063/1.321602
    
}

}

@article{Itano_1995,
	doi = {10.1088/0031-8949/1995/t59/013},
	url = {https://doi.org/10.1088/0031-8949/1995/t59/013},
	year = 1995,
	month = {jan},
	publisher = {{IOP} Publishing},
	volume = {T59},
	pages = {106--120},
	author = {Wayne M Itano and J C Bergquist and J J Bollinger and D J Wineland},
	title = {Cooling methods in ion traps},
	journal = {Physica Scripta},
	abstract = {In many experiments utilizing ion traps, the ions must be cooled in order to increase the precision and accuracy of the measurements. Laser cooling is very effective when it can be applied, but it can only be used with a few kinds of ions, since it depends critically on the details of the electronic level structure. Other methods, such as resistive cooling, active-feedback cooling, collisional cooling, radiofrequency side-band cooling, or sympathetic laser cooling, can be applied to many kinds of ions. Progress in cooling of trapped ions has made possible improved measurements of mass ratios and atomic spectra and the observation of new phenomena, such as the formation of ordered Coulomb "crystals" of ions.}
}

@MISC{Schrieffer1966-gm,
  author  = "Schrieffer, J R and Wolff, P A",
  journal = "Physical Review",
  volume  =  149,
  number  =  2,
  pages   = "491--492",
  year    =  1966
}

@ARTICLE{Dubielzig2021-qh,
  author   = "Dubielzig, T and Halama, S and Hahn, H and Zarantonello, G and
              Niemann, M and Bautista-Salvador, A and Ospelkaus, C",
  abstract = "We describe the design, commissioning, and operation of an
              ultra-low-vibration closed-cycle cryogenic ion trap apparatus.
              One hundred lines for low-frequency signals and eight
              microwave/radio frequency coaxial feed-lines offer the
              possibility of implementing a small-scale ion-trap quantum
              processor or simulator. With all supply cables attached, more
              than 1.3 W of cooling power at 5 K is still available for
              absorbing energy from electrical pulses introduced to control
              ions. The trap itself is isolated from vibrations induced by the
              cold head using a helium exchange gas interface. The performance
              of the vibration isolation system has been characterized using a
              Michelson interferometer, finding residual vibration amplitudes
              on the order of 10 nm rms. Trapping of 9Be+ ions has been
              demonstrated using a combination of laser ablation and
              photoionization.",
  journal  = "Rev. Sci. Instrum.",
  volume   =  92,
  number   =  4,
  pages    = "043201",
  month    =  apr,
  year     =  2021,
  language = "en"
}

@ARTICLE{Vrijsen2019-xr,
  author   = "Vrijsen, Geert and Aikyo, Yuhi and Spivey, Robert F and Inlek, I
              Volkan and Kim, Jungsang",
  abstract = "We report a highly efficient loading of 174Yb+ ions in a surface
              electrode ion trap by using single pulses from a Q-switched
              Nd:YAG laser to ablate neutral atoms, combined with a two-photon
              photo-ionization process. The method is three orders of magnitude
              faster to load a single ion as compared to traditional
              resistively heated sources and can load large collections of ions
              in seconds. The negligible thermal load of this method enables
              the use of this ablation-based loading scheme in ion traps
              operating under cryogenic conditions.",
  journal  = "Opt. Express",
  volume   =  27,
  number   =  23,
  pages    = "33907--33914",
  month    =  nov,
  year     =  2019,
  language = "en"
}

@ARTICLE{Shao2018-ak,
  author    = "Shao, H and Wang, M and Zeng, M and Guan, H and Gao, K",
  abstract  = "In this paper, we present a detailed method for rapidly and
               effectively loading a single 40Ca+ ion into a miniature linear
               Paul trap. Calcium atoms are generated by laser ablation, and
               the single ion is loaded and specific isotopes are selected by a
               two-step photo-ionization method. Compared to the traditional
               photo-ionization method in which atoms are typically emitted
               from a resistively heated oven, the advantages of laser ablation
               are that it can be precisely controlled, it can effectively
               avoid the problem of fluxed calcium deposition on the trap
               electrodes, and it restricts the generation of dark ions, which
               can influence atomic-precision spectroscopy measurements. With
               its short loading time, it is possible to load an ion in few
               seconds with the appropriate parameters. The technique of laser
               ablation has been applied to generate several species of atoms
               and ions at present, it will potentially benefit in the studies
               of optical frequency standards, quantum simulation and quantum
               information processing in an ultra-high vacuum condition and
               cryogenic system.",
  journal   = "J. Phys. Commun.",
  publisher = "IOP Publishing",
  volume    =  2,
  number    =  9,
  pages     = "095019",
  month     =  sep,
  year      =  2018,
  language  = "en"
}

@ARTICLE{Osada2022-oh,
  author    = "Osada, Alto and Noguchi, Atsushi",
  abstract  = "Trapped-ion quantum technologies have been developed for decades
               toward applications such as precision measurement, quantum
               communication and quantum computation. Coherent manipulation of
               ions' oscillatory motions in an ion trap is important for
               quantum information processing by ions, however, unwanted
               decoherence caused by fluctuating electric-field environment
               often hinders stable and high-fidelity operations. One way to
               avoid this is to adopt pulsed laser ablation for ion loading, a
               loading method with significantly reduced pollution and heat
               production. Despite the usefulness of the ablation loading such
               as the compatibility with cryogenic environment, randomness of
               the number of loaded ions is still problematic in realistic
               applications where definite number of ions are preferably loaded
               with high probability. In this paper, we demonstrate an
               efficient loading of a single strontium ion into a surface
               electrode trap generated by laser ablation and successive
               photoionization. The probability of single-ion loading into a
               surface electrode trap is measured to be 82\%, and such a
               deterministic single-ion loading allows for loading ions into
               the trap one-by-one. Our results open up a way to develop more
               functional ion-trap quantum devices by the clean, stable, and
               deterministic ion loading.",
  journal   = "J. Phys. Commun.",
  publisher = "IOP Publishing",
  volume    =  6,
  number    =  1,
  pages     = "015007",
  month     =  jan,
  year      =  2022,
  language  = "en"
}

@ARTICLE{Leibrandt2007-au,
  author    = "Leibrandt, David R and Clark, Robert J and Labaziewicz, Jaroslaw
               and Antohi, Paul and Bakr, Waseem and Brown, Kenneth R and
               Chuang, Isaac L",
  journal   = "Phys. Rev. A",
  publisher = "American Physical Society",
  volume    =  76,
  number    =  5,
  pages     = "055403",
  month     =  nov,
  year      =  2007
}

@ARTICLE{Leibrandt2009-ij,
  author    = "Leibrandt, D R and Labaziewicz, J and Clark, R J and Chuang, I L
               and Epstein, R J and Ospelkaus, C and Wesenberg, J H and
               Bollinger, J J and Leibfried, D and Wineland, D J and Stick, D
               and Sterk, J and Monroe, C and Pai, C-S and Low, Y and Frahm, R
               and Slusher, R E",
  abstract  = "A scalable, multiplexed ion trap for quantum information
               processing is fabricated and tested. The trap design and
               fabrication process are optimized for scalability to small trap
               size and large numbers of interconnected traps, and for
               integration of control electronics and optics. Multiple traps
               with similar designs are tested with 111Cd+, 25Mg+, and 88Sr+
               ions at room temperature and with 88Sr+ at 6 K, with respective
               ion lifetimes of 90 s, 300 $\pm$ 30 s, 56 $\pm$ 6 s, and 4.5
               $\pm$ 1.1 hours. The motional heating rate for 25Mg+ at room
               temperature and a trap frequency of 1.6 MHz is measured to be 7
               $\pm$ 3 quanta per millisecond. For 88Sr+ at 6 K and 540 kHz the
               heating rate is measured to be 220 $\pm$ 30 quanta per second.",
  journal   = "Quantum Inf. Comput.",
  publisher = "Rinton Press, Incorporated",
  volume    =  9,
  number    =  11,
  pages     = "901--919",
  month     =  nov,
  year      =  2009,
  address   = "Paramus, NJ"
}

@ARTICLE{Noguchi2020-sd,
  author   = "Noguchi, Atsushi and Yamazaki, Rekishu and Tabuchi, Yutaka and
              Nakamura, Yasunobu",
  abstract = "Electromagnetic fields carry momentum, which upon reflection on
              matter gives rise to the radiation pressure of photons. The
              radiation pressure has recently been utilized in cavity
              optomechanics for controlling mechanical motions of macroscopic
              objects at the quantum limit. However, because of the weakness of
              the interaction, attempts so far had to use a strong coherent
              drive to reach the quantum limit. Therefore, the single-photon
              quantum regime, where even the presence of a totally off-resonant
              single photon alters the quantum state of the mechanical mode
              significantly, is one of the next milestones in cavity
              optomechanics. Here we demonstrate an artificial realization of
              the radiation pressure of microwave photons acting on phonons in
              a surface acoustic wave resonator. The order-of-magnitude
              enhancement of the interaction strength originates in the
              well-tailored, strong, second-order nonlinearity of a
              superconducting Josephson junction circuit. The synthetic
              radiation pressure interaction adds a key element to the quantum
              optomechanical toolbox and can be applied to quantum information
              interfaces between electromagnetic and mechanical degrees of
              freedom.",
  journal  = "Nat. Commun.",
  volume   =  11,
  number   =  1,
  pages    = "1183",
  month    =  mar,
  year     =  2020,
  language = "en"
}

\appendix
\section{Calcium titanate as a long-lived ablation target}

As stated in Sec.~\ref{ablation}, pulsed laser ablation is relevant for loading ions and electrons into the Paul trap, however, the lifetime of the ablation target is crucial for the vacuum-sealed experiment running for months long.  The lifetime, or how many ablation pulses the target can endure, can vary over target materials~\cite{Leibrandt2007-au} and we here investigate such property of metallic calcium and ceramic calcium titanate as target materials and compare the results.

The 1064~nm-wavelength, nanosecond pulsed laser with pulse fluence of 3~J/cm$^2$ is shoot on the samples and the fluorescence of the calcium atoms can be observed when 423~nm-wavelength laser path crosses the atom jet. Figure~\ref{FigA} shows the observed fluorescence counts as the number of the applied ablations pulsed at the same target point increases. As can be seen in the plots, the signal for the calcium titanate (red) persists for thousands of ablation pulses in contrast to the metallic calcium (gray), meaning that the calcium titanate is a better target material than metallic calcium.

\begin{figure}[h]
\includegraphics[width=8cm]{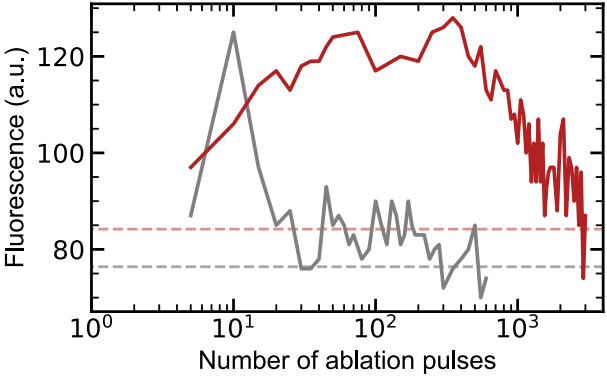}
\caption{\label{FigA} Fluorescence counts as a function of the number of ablation pulses. Red (gray) plot is for the calcium titanate (metallic calcium) with the background noise level indicated by a red (gray) broken line.}
\end{figure}

\end{document}